\documentclass[preprint]{elsarticle}
\usepackage[utf8]{inputenc}

\usepackage[superscript,biblabel]{cite}
\usepackage{geometry}
\usepackage{amssymb}
\usepackage{appendix}
\usepackage{amsmath}
\usepackage{bm}
\usepackage{float}
\usepackage{graphicx}
\usepackage{yhmath}
\usepackage{siunitx}
\usepackage{upgreek}
\usepackage{multicol}

\usepackage{natbib}

\usepackage{hyperref}
\hypersetup{colorlinks = true, linkcolor = blue, filecolor = magenta, urlcolor = cyan}

\usepackage[dvipsnames]{xcolor}

\begin{document}
\begin{frontmatter}
\title{In Situ 3D Spatiotemporal Measurement of Soluble Biomarkers in Organoid Culture}
\author[add1,add2]{Alexander~McGhee\corref{cor}}
\author[nrl,add2]{Eric~McGhee}
\author[add1,add2]{Jack~E~Famiglietti}
\author[add1,add2]{W~Gregory~Sawyer}
\cortext[cor]{Corresponding author; amcghee2@wisc.edu} 
\address[nrl]{Biomolecular Science \& Engineering Division Naval Research Laboratory Washington, DC 20375}
\address[aur]{Research and Development Aurita Bioscience Gainesville, FL 32601}
\address[add1]{Department of Mechanical Engineering, University of Wisconsin-Madison, Madison, WI 53706, USA}
\address[add2]{Department of Mechanical \& Aerospace Engineering Herbert Wertheim College of Engineering University of Florida Gainesville, FL 32601}
\begin{abstract}
Advanced cell culture techniques such as 3D bio-printing and hydrogel-based cell embedding techniques harbor many new and exciting opportunities to study cells in environments that closely recapitulate in-vivo conditions. Researchers often study these environments using fluorescence microscopy to visualize the protein association with objects such as cells within the 3D environment, yet quantification of concentration profiles in the microenvironment has remained elusive. Here, we present a method to continuously measure the time-dependent concentration gradient of various biomarkers within a 3D cell culture assay using bead-based immunoassays to sequester and concentrate the fluorescence intensity of these tagged proteins. This assay allows for near real-time in situ biomarker detection and enables spatiotemporal quantification of biomarker concentration. Snapshots of concentration profiles can be taken, or time series analysis can be performed enabling time-varying biomarker production estimation. Example assays utilize an osteosacroma tumoroid as a case study for a quantitative single-plexed gel encapsulated assay, and a qualitative multi-plexed 3D bioprinted assay. In both cases, a time-varying cytokine concentration gradient is measured. An estimation for the production rate of the IL-8 cytokine per second per osteosarcoma cell results from fitting an analytical function for continuous point source diffusion to the measured concentration gradient and reveals that each cell produces approximately two IL-8 cytokines per second. Proper calibration and use of this assay is exhaustively explored for the case of diffusion-limited Langmuir kinetics of a spherical adsorber. 
\end{abstract}
\journal{arxiv}
\end{frontmatter}
\begin{multicols}{2}
\section{Introduction}
\label{S:I}
The tumor microenvironment creates a protective immunosuppressive barrier and represents a significant obstacle to therapies \cite{Discher2009-ph,Hanahan2011-gn,Tabassum2015-ed,Vidi2013-vn}. Local measurements of the secreted cytokines and signaling molecules involved in immunoregulation within this interphase between tumor tissue and the non-cancerous surrounding tissues is particularly challenging due to difficulties in probing this exquisitely small volume. On cellular dimensions ( O(10) $\upmu$m in thickness) this total volume of the interphase region can be as small as a microliter (depending on the tumor size); in vitro interphase volumes further exacerbate the difficulty of measurement since the tumor explants and tumoroids are often much smaller than in vivo with interphase volumes of O(1-10)nL. The second major challenge with local measurements is that the in vitro tumor models are increasingly becoming three-dimensional, which introduces significant sampling and imaging challenges \cite{Drost2018-rj,LeSavage2022-wf}. Accurate measurements of local concentrations over nano-liter volumes in 3D pose significant challenges in both collection and detection. To overcome these challenges we have suspended ELISA beads within traditional matrigel-based 3D culture as well as 3D bioprinted Liquid Like Solid culture. These suspended ELISA beads are then able to participate in capture dynamics and when imaged in a standard confocal microscope, allow for quantifiable in-situ detection of cytokines in the immediate tumor vicinity. In this manuscript, we will discuss the approach of using fields of ELISA beads to measure biometric data such as spatiotemporal concentration gradients and cellular production rates.

Standard benchtop in vitro experiments enjoy a plethora of techniques to quantify biomarker concentrations\cite{Cox2019-ow}. Often centered around media samples captured during cell feeding, the majority of immunoassays offer robust measurement of biomarkers for time-point analysis. Although these benchtop immunoassays are the gold standard for many biomarkers, they typically do not have a sensitivity high enough for biomarkers that are scarcely produced and can only measure the average biomarker concentration in the entire assay volume. To overcome this issue, researchers have developed an in-situ immunoassay called FluoroSpot which enables the simultaneous measurement of multiple biomarkers within 2D cell culture at high sensitivity \cite{Janetzki2014-yq}. In this assay, the cells act as a point source for the diffusion of biomarkers, thus, a gradient of captured biomarkers surrounds each cell; this gradient is detected through the use of a fluorophore-conjugated antibody which binds to the captured biomarkers in a fashion commonly referred to as the sandwich ELISA \cite{Aydin2015-sz}. A fluorescence measurement device can then be used to measure the resulting intensity gradient and further analyzed to reveal the concentration gradient profile. Although the FluoroSpot method works well for 2D cell culture, there is currently no available analog for 3D culture.

Recent advancements in immunoassay development have enabled the detection of multiple biomarkers from a single sample by utilizing antibody-conjugated microbeads, commonly known as bead-based multiplexed ELISA assays. These assays are becoming increasingly popular with many off-the-shelf biomarker detection kits available from numerous manufacturers. Most of these assays utilize the sandwich ELISA method to capture and detect biomarkers in a solution through the use of proprietary flow cytometers such as Magpix®, LUNARIS™, and Quanterix® \cite{Djoba_Siawaya2008-mu}. Manufacturers of bead-based assays typically require the use of custom-designed flow cytometers for the detection of fluorescence intensity with their bead-based immunoassay to maximize the fluorescence intensity signal and achieve high sensitivity and linearity of the assay. Although these detection systems are custom-built, they all utilize the same basic principles outlined in Fulton et al\cite{Fulton1997-wh}. Briefly, microbeads with one or more unique fluorescence signatures are used to identify the biomarker specificity of each bead while other fluorescence signatures are used to detect biomarker concentration through a calibration curve transfer function\cite{Morgan2004-dj}. Great care is taken in the design of the bead adsorption kinetics to optimize and further enhance the detectability of the biomarkers\cite{Huovinen2021-sd}.  

\begin{figure}[t!]
\centering
\includegraphics[width = 1 \textwidth]{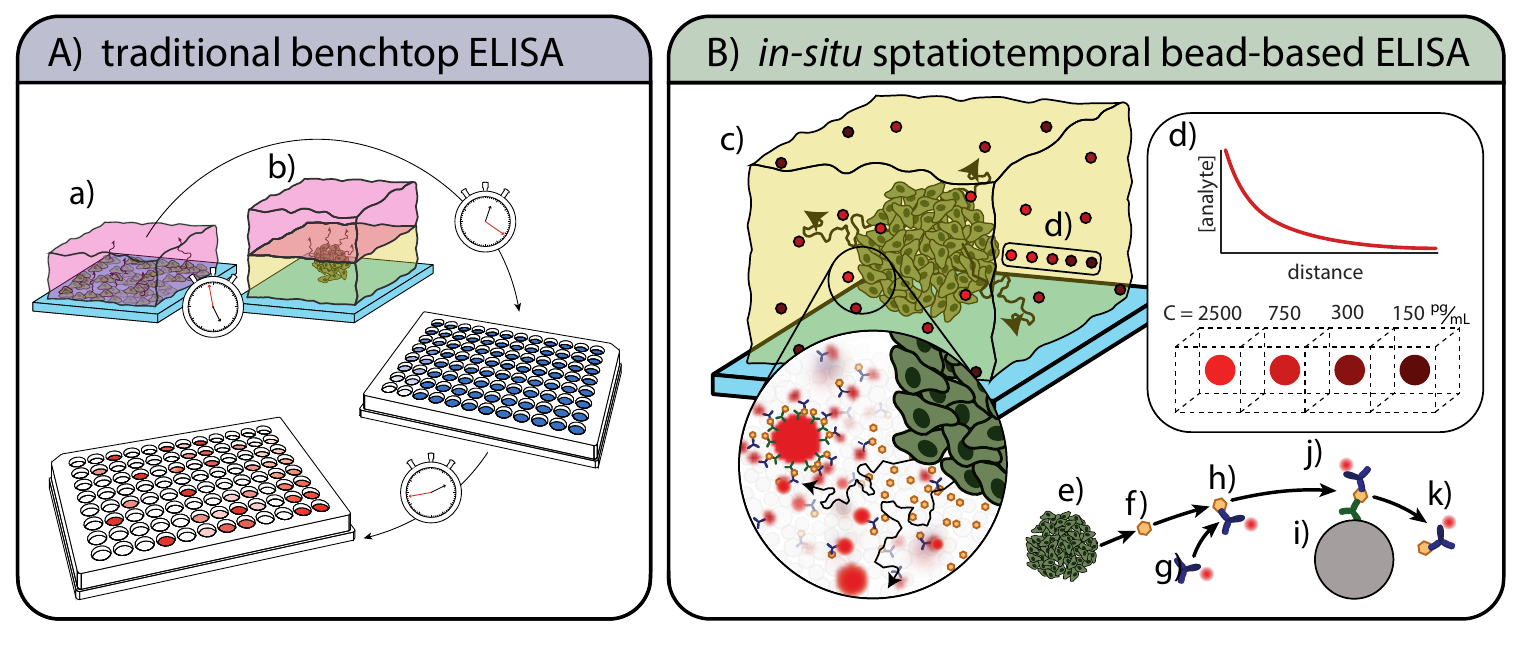}
\caption{Overview of traditional ELISA methods versus our approach A) typical bench-top methods for measuring biomarkers with immunoassays from (a) 2D cell culture and (b) 3D cell culture rely on the diffusion of biomarkers into a bath of growth media which is collected at various time points. These methods dilute the biomarker concentration by the total volume of growth media in the bath. Furthermore, the sample collection and testing can often take multiple days to obtain results. B) our method pictured in (c) utilizes immunosorbent beads scattered throughout the volume of a 3D cell culture assay. The bead fluorescence intensity corresponds to the local concentration of biomarker within a local bead volume of approximately 20 nL and these local concentrations may decrease with increasing distance from a source following a concentration gradient of a continuous point source (d). For a continuous source of biomarker from a collection of cells (e), each biomarker (f) released from a cell will bind to a detection antibody (g) to form a pair (h), this antibody biomarker pair then adsorbs to the immunosorbant bead (i) via conjugated capture antibodies (j). The antibody biomarker pair will desorb (k) from the capture antibody at some reverse reaction rate $k_r$. Through this process, the capture beads will produce a time-varying signal proportional to the local concentration of biomarker at any time. } \label{fig:CMP}
\end{figure}

Here, we utilize these bead-based immunoassays to develop an in-situ biomarker detection method for 3D cell culture. This technique uses fluorescence microscopy to detect fluorescence intensities corresponding to local biomarker concentration, as overviewed in (Fig.\,\ref{fig:CMP}(B)). Using this method, we are able to quantify the biomarker concentration in a singleplex assay and show that this technique is able to capture in situ spatiotemporal information of biomarker concentration in near real-time. Qualitative detection of biomarker production is also described in a multiplex assay using a much-simplified version of the quantitative method. Adaptation of the quantitative method for use in such a multiplex assay is straightforward but not presented. Furthermore, protocols for incorporating capture beads into standard 3D culture systems are developed for liquid-like solid 3D bioprinting \cite{Bhattacharjee2015-aw} and typical gel-based 3D culture \cite{Ravi2015-ln}. These results show the first successful implementation of a commercially available real-time in-situ immunoassay for 3D cell culture and 3D bioprinting assays in both quantitative singleplex, and qualitative multiplex fashions.
\end{multicols}
\begin{multicols}{2}
\section{Assay Creation} 
\label{S:AC}
\subsection{Building a Qualitative Biomarker Detection Assay}
\label{ss:BQualA}
The simplest form of this assay can be implemented in a qualitative manner with relative ease resulting in spatial information of biomarker presence within detectable ranges. In practice, a combination of cytometric beads and detection antibody within a 3D assay can provide a boolean interpretation of biomarker presence. In short, if a mean fluorescence intensity ($MFI$) exists in the proper channel after background subtraction, then the biomarker is present at detectable levels, and vice versa. However, the authors suggest careful consideration of Appendix B before use of a qualitative assay due to various issues related to the kinetics of singlepot-ELISA. For example, the use of custom cytometric beads or overly dense bead concentrations may lead to poor or misleading results. Nevertheless, this level of the assay can allow a user to rapidly probe an experiment to provide real-time spatial information about the state of the organoid in a 3D culture system.

\subsection{Building a Quantitative Biomarker Detection Assay}
\label{ss:BQuanA}
In this manuscript, we utilize cytometric beads in a manner that is far from the manufacturer's intended use. For example, measurement of the fluorescence intensity is achieved using a fluorescence microscope which has many associated issues with capturing quantitative information\cite{Fricker2006-cz,Jonkman2020-px,Royer2019-xt}. These issues ultimately lead to relatively large uncertainties of measured biomarker concentration, however, the ability to capture the spatiotemporal concentration of biomarker in real-time remains an attractive feature. 
The adaptation of these commercial assays to capture in-situ quantitative information for 3D culture systems demands attention to two critical considerations: the conversion of $MFI$ to biomarker concentration and determination of acceptable bead density. Typical bead-based assays assume a pure convection boundary at the bead surface coupled with an abundance of biomarker much beyond the total capture capacity of the beads in solution. These assumptions cannot be made upon adaptation to 3D culture systems which typically have minimal or no convection\cite{Nguyen2022-lh}. This leads to diffusion-limited reactions in the 3D environment and thus special considerations must be taken to ensure that measured bead $MFI$ is calibrated correctly. For this calibration, a standard curve is generated under pure diffusion, and we introduce a dynamic measurement based on the concentration gradient of the biomarker with known boundary conditions. The dynamic measurement is based on a flat wall diffusion gradient of biomarker from a fully mixed convective chamber into a 3D culture analog. The purpose of the diffusion chamber is to ensure the density of capture beads used in the assay is appropriate to the assay kinetics. Since each capture bead sequesters some amount of biomarker from the local environment, capture beads may compete with one another if they are too densely packed, effectively lowering the detected local concentration as shown in Fig.\,\ref{fig:CAL}(D). Thus, the capture bead concentration must be experimentally verified to ensure an appropriate density exists.
\section{Methodology in Demonstration} 
\label{S:M}
\subsection{Culture of osteosarcoma spheroids}
\label{ss:OC}
OS521 was cultured in a T-75 flask (Corning) using a complete culture medium with 0.3 mg/mL G418 (Mediatech) at $37^\circ$C, 5\% $CO_2$ \cite{Levings2009-fg}. After reaching confluence, cells were released from the flask using a trypsin dissociation reagent (TrypLE, Gibco) and resuspended in culture media to a density of approximately 106 cells/mL. This cell suspension was then distributed to an ultra-low attachment round-bottom 96-well plate (Corning ) and incubated at $37^\circ$C, 5\% $CO_2$ for 72 hours with daily media changes until tight spheroids were formed. Immediately before use in assays, spheroids were dyed using CellTracker Green CMFDA dye using standard manufacturer-specified protocols prior to the experiment.

\subsection{preparation of gel-based distributed bead assay}
\label{ss:GB}
A human IL-8 cytometric bead array (CBA) kit (BD biosciences) consisting of three components, IL8 capture beads, fluorophore-conjugated detection antibody, and lyophilized human IL8 standard was used in a matrigel based culture of osteosarcoma spheroids. First, the detection antibody was diluted in a solution of culture media as well as a solution of matrigel at a ratio of 1:200. This ratio may vary between assays and manufacturers since the detection antibody concentration is not consistent. For this assay, the target detection antibody concentration allows for 10x the number concentration of detection antibodies to the highest local number concentration of biomarkers expected. The detection antibody concentration can be determined using the protocol outlined in \ref{ss:DAC}. Next, the capture beads were added to the matrigel solution at a concentration of 50 beads/$\upmu$L (the concentration of capture beads given in BD CBA kits is approximately 3,000 beads/$\upmu$L), gently pipette mixed, and placed on ice. An osteosarcoma spheroid, chilled on ice, was then added to the matrigel-bead solution and gently pipette mixed. The solution was then pipetted onto a glass-bottom well to form a dome, the spheroid was gently moved towards the center, and the sample was placed in an incubation chamber at $37^\circ$C, 5\% $CO_2$ until the matrigel was set (approximately 2 minutes). Finally, the media + detection antibody solution was added to the culture dish ensuring the matrigel was fully submerged. The sample was then placed on a fluorescent microscope and imaged every 4 hours using laser channels of: GFP, to capture the OS521 spheroid; cy5, to capture bead positions; and TRITC, to capture detection antibody intensity. The resulting flourescent images can be seen in Fig.\,\ref{fig:SMA}(A).
\subsection{preparation of 3D printed bead assay}
\label{ss:3DP}
Four BD CBA kits of human IL-11, human IL-8, human IL-6, and human VEGF are checked for cross-reactivity by consulting the BD CBA manual. A solution of LLS is created using culture media according to the protocol outlined in Bhattacharjee et al.\cite{Bhattacharjee2016-dm}. Detection antibodies from each kit are diluted into culture media as well as a solution of LLS at a ratio of 1:200 v/v. Capture beads from each kit are combined with the LLS solution in separate aliquots at a concentration of 50 beads/$\upmu$L. In a separate aliquot, an osteosarcoma spheroid is combined with a small amount of the LLS solution and printed into a glass-bottom petri dish filled with LLS following a method outlined  in Bhattacharjee et al.\cite{Bhattacharjee2016-dm}. Each capture bead solution is then printed in a 4-arm configuration centered about the osteosarcoma spheroid as shown in Fig.\,\ref{fig:SMA}(C). Finally, the media + detection antibody solution is carefully added atop the printed structures being careful not to disturb the sample. The sample is then placed on a fluorescent microscope and imaged every 4 hours using both the bead detection (cy5) channel as well as the detection antibody (TRITC) channels. 
\section{Results}
\label{S:R}
\subsection{Calibration}
\label{ss:Cal}
\begin{figure}[t!]
\centering
\includegraphics[width = 0.9 \textwidth]{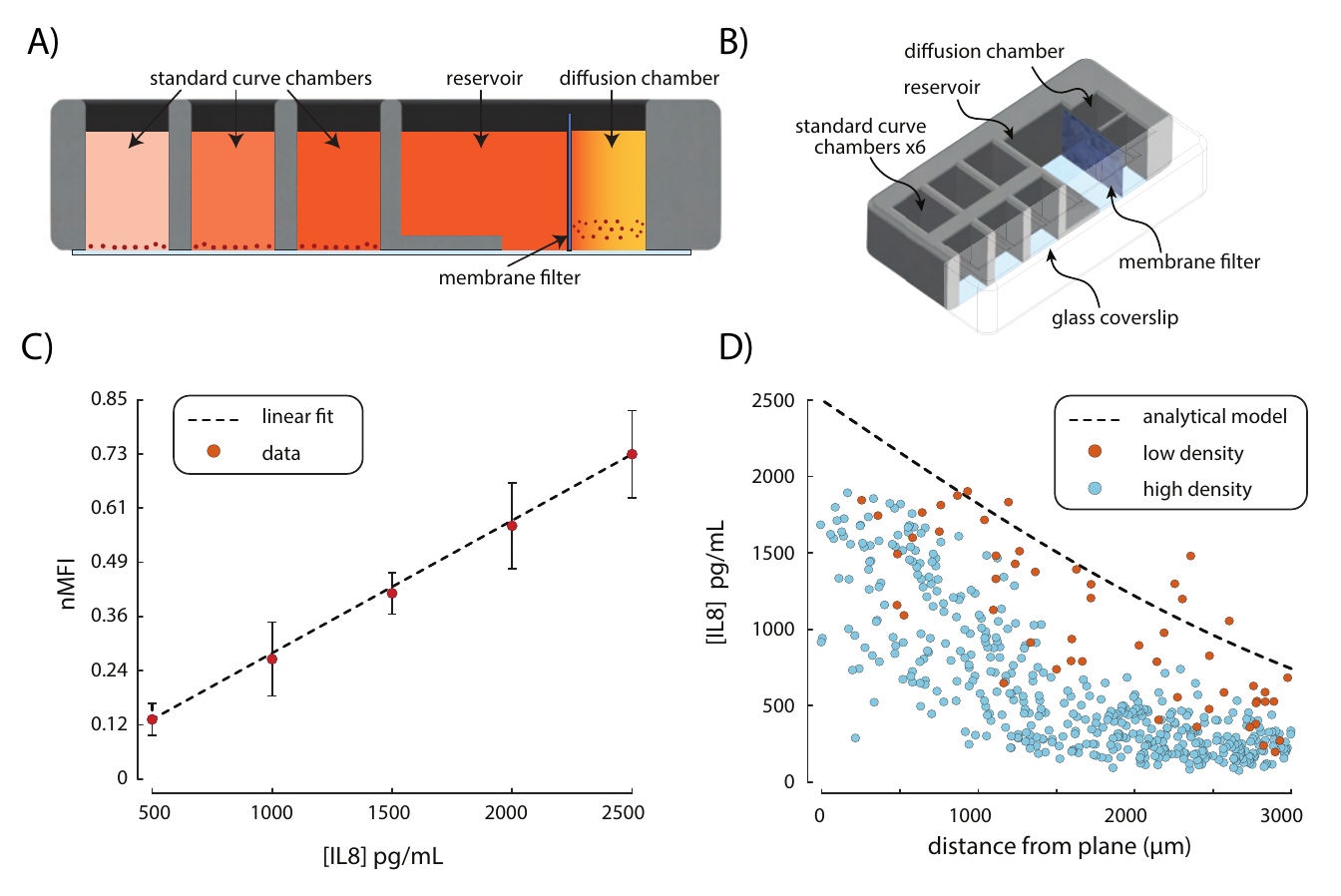}
\caption{Calibration device for standard curve and dynamic diffusion curve analysis. A) a cross-sectional view of the calibration device showing the standard curve wells, reservoir, diffusion chamber, and membrane filter. The colorscale from light pink to dark orange signifies the concentration of biomarkers in from low to high respectively. B) an isometric view of the calibration device shows the overall construction. C) The standard curve is generated using a linear spaced concentration of biomarker from 500 to 2500 pg/mL which corresponds to this assays linear range. The $MFI$ of each bead is measured and then normalized ($nMFI$) by the top of the dynamic range. D) Beads in the diffusion chamber are measured after 10 hours and the $nMFI$ is transferred into cytokine concentration using the standard curve. An analytical model for flat wall diffusion (black dashed line) is compared to two bead densities $\Gamma = 0.1$ (low density, red circles) and $\Gamma =1$ (high density, blue circles). These results show that as $\Gamma$ becomes much less than one, the beads measurement more closely follows the expected value. } 
\label{fig:CAL}
\end{figure}

We developed a PDMS device for the calibration of $MFI$ versus the biomarker concentration using a serial dilution of biomarker standard as well as the validation of bead kinetics. For this study, we used the human IL-8 CBA kit to produce a calibration curve and used the results to produce a concentration gradient profile for two bead concentrations - 50 and 500 beads/$\upmu$L. The results presented in Fig.\,\ref{fig:CAL}(D) show that the bead concentration of 50 beads/$\upmu$L produces a concentration profile similar to the analytical model described by eq.\ref{eq:FWD} and the bead concentration of 500 beads/$\upmu$L greatly underpredicts the analytical model. Kinematic variables of the assay were determined using the protocols described in \ref{S:K} finding the antibody + biomarker diffusion coefficient $D$ to be 55 $\upmu$ $m^2$/s, the reverse reaction rate $k_r$ to be approximately 0.001 $1/s$, and the number of capture sites per bead N to be 2.6 E6 sites. Using these kinematic variables with a biomarker concentration of the top standard $C$ = 2500 pg/mL, the equilibrium bead coverage $\theta_{eq}$  was found to be 0.12. Using these variables in eq.\ref{eq:Gamma} reveals the sequestration ratio $\Gamma$ for bead concentrations of 50 beads/$\upmu$L and 500 beads/$\upmu$L to be 0.1 and 1 respectively. This result demonstrates that eq.\ref{eq:Gamma} can accurately predict the appropriate bead density when the sequestration ratio is $\Gamma << 1$.
\end{multicols}
\begin{figure}[t!]
\centering
\includegraphics[width = 0.86 \textwidth]{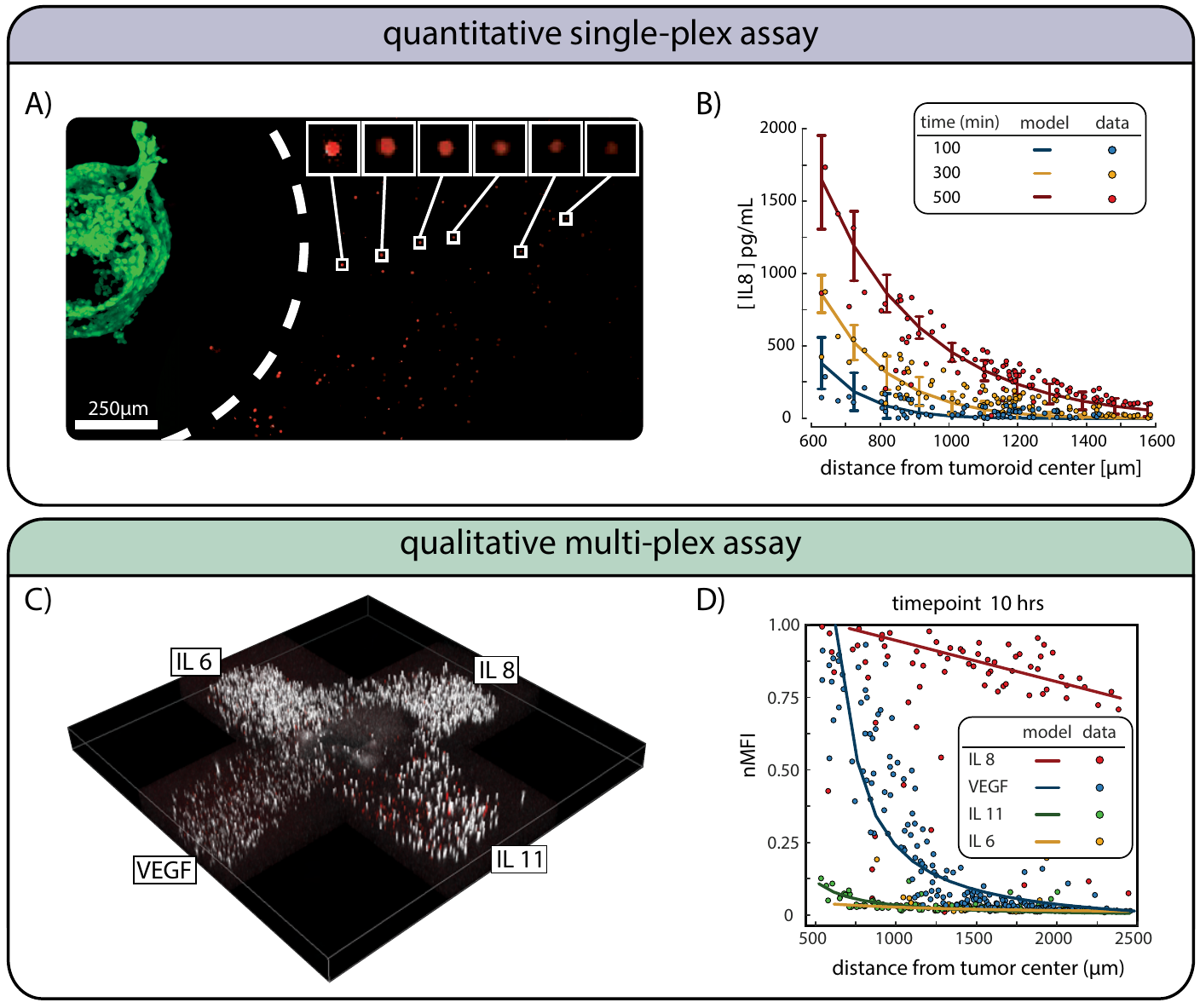}
\caption{Quantitative single-plex, and qualitative multi-plexed assay results for an osteosarcoma spheroid. A) a max intensity z-projection shows a 500 $\upmu$m diameter osteosarcoma spheroid in green on the left with a white dashed line corresponding to the area where no detection beads are found. Breakouts of six beads at various radial distances from the spheroid show a decreasing intensity with increasing radial distance. B) The resulting transformation of the measured bead intensities to IL8 concentration for timepoints 100, 300, and 500 minutes shows good agreement with the analytical model for diffusion from a continuous point source with a production rate $q$ of approximately 2 IL-8 proteins per cell per second. C) A 3D printed bead array surrounding an osteosarcoma tumoroid shows the relative position of four bead types IL6, IL8, IL11, and VEGF. D) The detectability of each biomarker can be compared directly with IL8 and VEGF showing relatively high readings for each, and relatively low readings for IL-11 and IL-6. These readings would indicate, using the qualitative assay, that IL-8 and VEGF are present while IL-11 and IL-6 are not being produced in sufficient quantities to be measured. } \label{fig:SMA}
\end{figure}
\begin{multicols}{2}
\subsection{Singleplex matrigel-based assay}
\label{ss:SA}
OS521 is known to produce the IL-8 cytokine in abundance, so we chose to focus on this cytokine for our trial study\cite{Gross2018-ef}. After the IL-8 capture bead density and $MFI$ vs concentration calibration curve were found, we analyzed the resulting intensity gradient of the beads surrounding the spheroid to recover the concentration gradient for three time points 100, 300, and 500 minutes as shown in Fig.\,\ref{fig:SMA}(A,B). An analytical solution for a continuous point source given by Crank\cite{Crank1979-cg} in eq.\ref{eq:CPS} 
\begin{equation}
C\left ( r,t \right ) = \frac{q}{4\pi Dr}erfc\left [ \frac{r}{2\sqrt{Dt}} \right ] 
\label{eq:CPS} 
\end{equation} 
is fitted to the experimental data using the rate of production $q$ as a free variable with $r$ the radial distance from the point source as the independent variable. The diffusion coefficient for the detection antibody $+$ biomarker $D$, and the assay time in seconds $t$ are known constants for each curve analyzed. The rate of cytokine production $q$ for the 500$\upmu$m diameter spheroid (estimated to be $\sim$10k cells) was found to be nearly 2 IL-8 proteins per cell per second by fitting eq.\ref{eq:CPS} to the concentration gradient at multiple timepoints.
\subsection{Multiplex LLS-based assay }
\label{ss:MA}
Bead-based immunoassays are often used to measure multiple biomarkers from the same sample, and thus the detection and capture antibodies have been optimized to ensure little to no cross-reactivity. We utilize this capability in a 3D bioprinted assay by separating the bead types by location in a 4-arm configuration centered around the spheroid as shown in Fig.\,\ref{fig:SMA}(C). The bead intensity gradient for the IL-11, IL-8, IL-6, and VEGF was measured for multiple time points and presented in Fig.\,\ref{fig:SMA}(D). These results show that the four cytokines measured have different concentration gradient time profiles and thus four different production rates. No calibration was completed for this multiplexed assay so results remain qualitative.
\section{Discussion}
\label{S:D}
The rise of 3D culture provides exciting opportunities to study in vitro biological systems in their closest analogous configuration to in vivo\cite{Alhaque2018-ui,Jacks2002-ot,Lv2017-uh,Nguyen2022-lh}. These methods offer potentially enhanced biological complexity through mass transport modification, 3D intercellular contact, cell-matrix signaling, customizable geometric configuration of tissues, tissue shielding, and allow complex immune interaction. The addition of perfusion into these 3D culture systems allows for long-duration cultures providing opportunity for emergent biological complexity to flourish in vitro\cite{Balasubramanian2021-wv,Eiraku2011-si,Jackson2016-ib,Oshima2017-tq}. Such methods are of keen interest in the pharmaceutical industry as the development and deployment of in vivo-like high throughput drug screening may offer better drug development outcomes\cite{Hughes2011-lv,Lovitt2014-ye}.  Unfortunately, interrogating biochemical compositions of 3D in vitro systems using standard bioassays can be extremely challenging to near impossible. Therefore, new developments in bioassay technologies are needed to provide the insights researchers have come to know as standard for 2D in vitro systems.

In this study, we have examined the potential for the use of a commercially available bead-based immunosorbent assay as a viable method for the in-situ measurement of cytokine concentration within 3D culture assays. This assay method not only represents the first of its kind measurement for 3D culture systems but is also easily implemented using this manuscript as a guide. Additionally, we have shown that this assay can be implemented in simple gel-based culture systems as well as more advanced 3D bioprinting systems at low cost. Typical bead-based assays supply approximately 300,000 beads, and more than enough detection antibody to coat each bead at the highest standard. The 3D assays used in this study used a bead density of 50 beads/$\upmu$L with a total gel volume of approximately 100$\upmu$L per sample. At this ratio, this experiment can be repeated 60 times at a total cost of a single bead-based assay kit. At the time of this writing, an IL-8 CBA kit costs approximately \$278.00 making the cost of each experiment less than five dollars. If a 3D printing technique were implemented, as we have done in this manuscript, this cost can be reduced significantly. 

Bead-based immunoassays are not typically designed to be used in the manner we prescribe here. Some assays may not be viable due to incompatible assay kinetics, background saturation, cross-reactivity, etc. For example, the immunoassay must be capable of single-pot capture and detection which requires the antigen to have unique binding epitopes for the paratopes of the capture and detection antibodies. Many assays are capable of single-pot capture and detection but do not recommend this incubation method due to a reduction in assay sensitivity via the inclusion of detection antibodies within the sample which may obscure signals at the low end of the detection range. In our experience, these factors typically result in a 10 to 20\% reduction of the assays reported range when used in our method. However, since our assay assumes the pure diffusion of biomarkers from one or more point sources, the local concentration of biomarkers accumulates over time and may eventually become detectable. Finally, the active implementation of this assay may limit biological interaction. The inclusion of detection antibody and subsequent conjugation of the detection antibody in solution may affect intercellular chemokine signaling and as a result, change cellular behavior. This challenge could be circumvented by cycling of detection antibody for periods before imaging, especially in thin sections of 3D gel-based assays or through perfusion pulsation in LLS-based experimental platforms.

\section{Conclusions}
We developed a novel method for the utilization of commercially available bead-based immunosorbent assays for use in 3D cell culture platforms. Two 3D culture systems, a simple matrigel-based assay, and an advanced 3D bioprinting assay were studied using osteosarcoma spheroids to generate and measure multiple cytokines of interest. Protocols to back out the kinematic variables from commercial cytometric assays have been developed and allow users to easily calculate the ideal beads density and the detection antibody concentration appropriate for an individualized assay. Finally, we show that this assay can be easily implemented and adapted for use in multiple 3D culture systems in both qualitative and quantitative capacities.
\end{multicols}
\setcounter{section}{0}
\renewcommand{\thesection}{Appendix \Alph{section}}
\section{Bead Measurements}
\label{S:BM}
Commercial bead-based multiplex immunoassays use one or more fluorescent dyes incorporated into the bead to identify the antigen specificity within a mixture of many bead types\cite{Breen2015-ll,Holmes2007-sl}. This internal fluorescence can be used to locate the bead centers within a volumetric scan of the 3D assay. Each bead in the volumetric scan should be cropped into individual max intensity projections using the centroid of the bead as measured by the internal fluorescence channel. Once the bead has been located and cropped, each bead image should then be analyzed in the fluorescence channel of the respective detection antibody using the $MFI$ of a circular masked region with dimensions equal to the diameter of the bead in real-world units. The uncropped inverse space of the full volume can then be used to develop a background fluorescence intensity used for background subtraction. Positive fluorescence intensity for any bead can be interpreted as detection of the respective biomarker in that location. 
\section{Calibration of Quantitative Biomarker Detection Assay}
\label{S:CAL}
\subsection{Calibration curve generation }
\label{ss:CC}
This assay utilizes a bead fluorescence intensity measurement using a fluorescence microscope and assumes the intensity is proportional to the local concentration of biomarkers due to one or more point sources in a 3D cell culture assay. This relationship between bead intensity and local biomarker concentration must be experimentally calibrated. To accomplish this, we have created a PDMS chip featuring six wells, and a reservoir $+$ diffusion chamber which is a rectangular well separated by a filter membrane (as shown in Fig.\,\ref{fig:CAL}(A,B)). The six wells are used to create a standard curve by creating a serial dilution of the biomarker standard. Detection antibodies and beads are added to the standards and the beads are allowed to sink to the bottom of each well. The detection antibody associates with the biomarker in solution and the beads are allowed to reach an adsorption equilibrium for their respective concentrations in a diffusion-limited reaction. 

To measure the assay kinetics within a 3D culture system, the diffusion chamber is first filled with the 3D culture medium mixed with capture beads and detection antibody distributed in a manner described in sections \ref{ss:GB} and \ref{ss:3DP}. In the reservoir, a solution of the highest concentration standard mixed with detection antibodies is filled to the same level as the 3D culture medium. 
The sample is placed onto a fluorescence microscope and the beads within the standard wells are used to calibrate the microscope intensity settings such that the dynamic range is maximized for the lowest and highest bead intensities used in the standards. Once a dynamic range is established, the bead $MFI$ should be measured according to the protocol outlined above to establish the standard curve. The detection settings (laser power, gain, lens, zoom, etc.) should be recorded and kept constant for all future measurements referring to this standard calibration. The standard concentration versus measured bead $MFI$ should be analyzed according to the manufacturer's instructions to create a calibration curve. This calibration curve should then be used to transform the intensity measurements from the assay into biomarker concentration.

Beads within the diffusion chamber are measured every two hours until an intensity gradient fills the view. The resulting concentration profile is then compared to the analytical function for flat wall diffusion given by Crank\cite{Crank1979-cg} in eq.\ref{eq:FWD}. 
\begin{equation}
C\left ( x,t \right ) = C_o \; erfc\left [ \frac{x}{2\sqrt{Dt}} \right ] 
\label{eq:FWD} 
\end{equation}  
Where $C(x,t)$ is the time-varying concentration at some distance $x$ from the flat wall of constant concentration $C_0$ and $D$ is the diffusion coefficient for the biomarker + detection antibody. This diffusion coefficient can be measured using the protocol outlined in section \ref{ss:D}. If the measured concentration profile does not match the analytical function, the bead density must be decreased. Refer to section \ref{ss:IBD} for more information related to bead density.  
\subsection{Determination of ideal bead density }
\label{ss:IBD}
Langmuir kinetics can be used to approximate the sequestration of biomarkers by each bead within the 3D assay with the number of sequestered biomarkers reaching an equilibrium surface concentration ($\theta_{eq}$ ) for some given local concentration ($C$) according to eq.\ref{eq:Teq}. 
\begin{equation}
\theta_{eq}= \frac{K_{eq}C}{1+K_{eq}C}
\label{eq:Teq} 
\end{equation} 
The equilibrium constant $K_{eq}$ describes the binding affinity of the capture antibody on the bead surface to the biomarker. This equilibrium constant is unique to each assay and can be determined using eq.\ref{eq:Keq} after the kinetic variables have been found (refer to \ref{S:K} for more information). To find the maximum bead density for each 3D assay, we consider the ratio of the total number of surface sites occupied at equilibrium ($q_{eq}$) versus the number of biomarkers ($q_V$) within some local volume ($V$) surrounding the bead for a given local concentration ($C$). This sequestration ratio can be found using eq.\ref{eq:Gamma}.
\begin{equation}
\Gamma= \frac{q_{eq}}{q_{v}} = \frac{N\theta_{eq}}{CV}
\label{eq:Gamma} 
\end{equation} 
Where $N$ is the number of capture sites on the bead, and $\theta_{eq}$  is the surface coverage at equilibrium. When the ratio given in eq.\ref{eq:Gamma} is much less than one, the sequestration of biomarkers from a bead does not significantly affect the local concentration of neighboring beads. In this case, the maximum bead density can be calculated using eq.\ref{eq:rho}.
\begin{equation}
\rho= \frac{1}{V}
\label{eq:rho} 
\end{equation} 
\section{Determination of Bead Assay Kinetic Variables }
\label{S:K}
Many of the variables used in \ref{S:CAL} are considered proprietary information by the manufacturer of the bead assay and thus are not readily available. However, these variables can be found through experimentation. The following sections outline the procedures for estimating the kinetic variables for any commercially available assay in a diffusion-limited system.
\subsection{Determination of detection antibody + biomarker diffusion coefficient }
\label{ss:D}
A solution of detection antibody mixed with a 1:1 ratio of biomarker is added to the gel solution used in the 3D culture platform then set within a glass-bottom culture dish. Fluorescence recovery after photobleaching (FRAPS) is then used to determine the diffusion coefficient according to the protocol outlined in Deschout et. al\cite{Deschout2010-kf}.
\subsection{Determination of the reaction rates}
\label{ss:RR}
The detection antibody and biomarker both attach and detach from the capture bead at a forward reaction rate $k_f$ and reverse reaction rate $k_r$. The forward reaction rate can be approximated using eq.\ref{eq:kf} adapted from the Einstein-Smoluchowski equation describing the diffusional association of two reactive species in solution\cite{Berg1985-wd,Smoluchowski1917-fz}. This forward reaction rate describes the total number of analyte impinging on the surface of the entire bead normalized by the number of capture sites and describes the rate the analyte will impinge on any given capture site. 
\begin{equation}
k_f =\frac{4\pi DR}{N}
\label{eq:kf} 
\end{equation} 
where $R$ is the radius of the capture bead, $N$ is the number of capture sites on the bead, and $D$ is the detection antibody + biomarker diffusion coefficient.
The reverse reaction rate should be determined experimentally since this rate is determined by the antibody-antigen binding affinity. Begin by transferring a bead that has reached equilibrium surface coverage of biomarker and detection antibody \footnote{Use beads from the highest concentration standard when measuring off-rate kinetics} to a buffer solution with a zero concentration of biomarker and detection antibody. The fluorescence intensity of the bead should then be measured every 5 minutes for 20 minutes. The resulting $MFI$ measured at some time $t$ can be used to find the reverse reaction rate $k_r$  using eq.\ref{eq:kr}.  
\begin{equation}
k_r= -\frac{1}{t}\; ln\left ( \frac{MFI - b}{MFI_0} \right )
\label{eq:kr} 
\end{equation}
Where $MFI_0$ is the initial mean fluorescence intensity and $b$ is the background fluorescence intensity. This exponential decay is predicted by the Langmuir adsorption model\cite{Langmuir1918-nx}.
The equilibrium constant $K_{eq}$ is simply the ratio of the forward and reverse reaction rate as seen in eq.\ref{eq:Keq} 
\begin{equation}
K_{eq}=\frac{k_f}{k_r} =\frac{4\pi DR}{N k_r}
\label{eq:Keq} 
\end{equation}
\subsection{Estimation of the number of capture sites $N$ and equilibrium surface coverage $\theta_{eq}$ }
\label{ss:N}
Kinematic variables for the Langmuir equilibrium surface coverage $\theta_{eq}$  versus local biomarker concentration $C$ can be found using the $MFI$ versus $C$ data provided by the standard curve. First, the biomarker concentration versus $MFI$ profile is cropped in the linear range. This allows for the assumption that the $MFI$ normalized by the maximum reported $MFI$ will be proportional to $\theta_{eq}$  normalized by the maximum equilibrium surface coverage at the highest biomarker concentration in the linear region max according to eq.\ref{eq:MFITeq}
\begin{equation}
\frac{MFI}{MFI_{max}} = \frac{\theta_{eq}}{\theta_{max}}
\label{eq:MFITeq} 
\end{equation}
Substituting eq.\ref{eq:Keq} into eq.\ref{eq:Teq} reveals the relationship between $\theta_{eq}$ and the assay kinematic variables for diffusion-limited reactions. 
\begin{equation}
\theta_{eq} = \frac{4\pi DRC}{Nk_r + 4\pi DRC}
\label{eq:Teq2} 
\end{equation}
Finally, substituting eq.\ref{eq:Teq2} into eq.\ref{eq:MFITeq} allows the reported $MFI$ vs concentration profile within its linear range to be used to fit eq.\ref{eq:MFITeq2} using the number of capture sites on each bead $N$ and $\theta_{max}$ as free variables. 
\begin{equation}
\frac{MFI}{MFI_{max}} = \frac{4\pi DRC}{\theta_{max}\left ( Nk_r + 4\pi DRC \right )} 
\label{eq:MFITeq2} 
\end{equation}
\subsection{Estimation of detection antibody concentration }
\label{ss:DAC}
The concentration of the detection antibody should be high enough to ensure all biomarkers generated within the 3D culture assay are captured by the detection antibody before reaching the embedded capture bead. However, the background fluorescence signal increases with increasing detection antibody concentration. We, therefore, utilize the calibration device to determine the detection antibody concentration in the following manner. The diffusion chamber from the calibration device is filled with a PBS 1X buffer solution and the detection antibody solution is added to one side of the chamber separated by the membrane filter. The fluorescence intensity of the second chamber is continuously measured every 5 minutes until an intensity gradient fills the image of a fluorescent microscope. Assuming the $MFI$ of each pixel is directly proportional to the antibody concentration, the flat wall diffusion equation from eq.\ref{eq:FWD} can be used to fit the intensity versus distance profile from the filter membrane using the concentration as a free variable. The diffusion coefficient of the detection antibody can be found using the FRAP method described in \ref{ss:D}. 
\section*{Declaration of competing interest}
 The authors declare that they have no known competing financial interests or personal relationships that could have appeared to influence the work reported in this paper.
\section*{Acknowledgements}
The authors would like to thank Dr. Parker Gibbs for generously donating OS521Oct4 cell lines to this work.
\bibliographystyle{unsrt}
\bibliography{Ref_paperpile}
\end{document}